\documentclass{elsart}

\usepackage{graphicx}
\usepackage{ifthen}
\usepackage{xspace}

\usepackage{amsmath}

\newcommand{\micron}{\ensuremath{\mu\mathrm{m}}\xspace}

\begin{document}

\begin{frontmatter}
\title{Study of Cabibbo Suppressed Decays of the $D^+_s$ Charmed-Strange Meson involving a $K^0_S$}

\date{\today}

\collaboration{The~FOCUS~Collaboration}\footnotemark
\author[ucd]{J.~M.~Link}
\author[ucd]{P.~M.~Yager}
\author[cbpf]{J.~C.~Anjos}
\author[cbpf]{I.~Bediaga}
\author[cbpf]{C.~Castromonte}
\author[cbpf]{A.~A.~Machado}
\author[cbpf]{J.~Magnin}
\author[cbpf]{A.~Massafferri}
\author[cbpf]{J.~M.~de~Miranda}
\author[cbpf]{I.~M.~Pepe}
% ...

\address[ucd]{University of California, Davis, CA 95616}
\address[cbpf]{Centro Brasileiro de Pesquisas F\'\i sicas, Rio de Janeiro, RJ, Brazil}
% ...

\footnotetext{See \textrm{http://www-focus.fnal.gov/authors.html} for additional author information.}

%\pacs{14.20.Lq,13.30Eg}

\begin{abstract}
We study
the decay of $D^+_s$ mesons into final states involving a $K_S^0$ and
report the discovery of Cabibbo suppressed decay modes 
$D_s^+\rightarrow K_S^0 \pi^-\pi^+\pi^+$ (179$\pm$36 events)
and $D_s^+\rightarrow K_S^0 \pi^+$ (113$\pm$26 events). The
branching ratios for the new modes are {$\frac{\Gamma(D_s^+\rightarrow
K_S^0\pi^-\pi^+\pi^+)}{\Gamma(D_s^+\rightarrow
K_S^0K^-\pi^+\pi^+)}$} = 0.18$\pm$0.04$\pm$0.05 and 
{$\frac{\Gamma(D_s^+\rightarrow K_S^0\pi^+)}{\Gamma(D_s^+\rightarrow
K_S^0K^+)}$} = 0.104$\pm$0.024$\pm$0.013.  
 
\end{abstract}
\end{frontmatter}

%\maketitle

%%% Body begins here

An essential ingredient to accurately model backgrounds
in heavy quark systems involves the identification and categorization  
of missing decay channels in the charm sector.
This is particularly important 
for the $D^+_s$ decays where a substantial part of its hadronic decay rate is yet
to be identified. Only two $D^+_s$ Cabibbo suppressed decays have been reported, 
namely $D^+_s \rightarrow K^+\pi^+\pi^-$~\cite{Link:2004f,Frabetti:1995e} and its resonance substructure
and  $D^+_s \rightarrow K^+K^+K^-$~\cite{Link:2002i}.  It was found that  
$\frac{D^+_s \rightarrow K^+\pi^+\pi^-}{D^+_s \rightarrow K^+K^-\pi^+}$ = 0.127 $\pm$ 0.007 $\pm$ 0.014 and 
$\frac{D^+_s \rightarrow K^+K^-K^+}{D^+_s \rightarrow K^+K^-\pi^+}$ = (8.95 $\pm$ 2.12 $^{+2.24}_{-2.31}) \times 10^{-3}$. The two Cabibbo suppressed channels differ by an 
order of magnitude (partly due to phase space) and additional decays are needed 
to establish patterns.  
In this paper we report the discovery of two Cabibbo suppressed decays of the 
$D^+_s$ meson; $D^+_s \rightarrow K_S^0\pi^-\pi^+\pi^+$ and $D^+_s \rightarrow K_S^0\pi^+$.
No inclusive estimates of the branching fraction for $D^+_s \rightarrow K_S^0\pi^-\pi^+\pi^+$ have been reported, but 
several predictions exist for the branching ratio of 
$\frac{D^+_s \rightarrow K_S^0\pi^+}{D^+_s \rightarrow K_S^0K^+}$~\cite{Verma:1991a,Buccella:1995a,Buccella:1996a}.  
Throughout this paper, 
charge conjugate modes are implied unless explicitly stated otherwise.

{\bf II. THE FOCUS EXPERIMENT}

The data come from 6 billion events recorded during the 1996-1997 fixed 
target run at Fermilab.
Electrons and positrons with an endpoint energy of approximately
300 GeV bremsstrahlung, yielding  photons which interact in a segmented
beryllium-oxide target to produce charmed particles. The average photon energy 
for events which satisfy our trigger is approximately 175~GeV. 
Charged particles are tracked and momentum analyzed by a system of silicon vertex 
detectors~\cite{Link:2002ts} in the target region, multi-wire proportional chambers
downstream of the interaction region, and two oppositely polarized dipole magnets.  
Particle identification is
performed by three threshold \v{C}erenkov counters, two electromagnetic calorimeters,
a hadronic calorimeter, and two muon systems. The main FOCUS trigger required tracks outside
of the central region and approximately 25~GeV (or more) of energy in the hadron calorimeter.

$D_s^+$ decays are reconstructed using a candidate driven
vertex algorithm~\cite{Frabetti:2001pg}. A decay vertex is formed from the reconstructed 
charged tracks. The $K_S^0 \rightarrow \pi^+\pi^-$ decays are reconstructed 
using techniques described
elsewhere~\cite{Link:2001dj}. Briefly, $K_S^0 \rightarrow \pi^+\pi^-$ decays can occur anywhere 
along the spectrometer. 
Depending on where the decays occur (upstream of the first magnet or inside the magnetic field)
and on how many multi-wire proportional chambers each pion passes, the $K^0_S$ are given a type 
number and the different types vary in mass resolution and in purity.
The momentum
information from the $K_S^0$ and the charged tracks is used to form a
candidate $D$ momentum vector, which is intersected with other tracks to find
the primary (production) vertex. Even though it is possible for the production 
vertex to be identified with a single track plus the $D^+_s$ momentum vector,
the signal quality is greatly improved by demanding at least two primary tracks. 
Events are selected based on several criteria. The
confidence level for the production vertex and for the charm decay vertex must be 
greater than 1$\%$. 
The likelihood for each charged particle to be a proton,
kaon, pion, or electron based on \v{C}erenkov particle identification is used to
make additional requirements~\cite{Link:2001pg}.  We define a $\chi^2$-like variable $W_i$ as 
$-2 \ln(\textrm{likelihood}_i)$ for the hypothesis $i$.
In order to reduce
background due to secondary interactions of particles from the production
vertex, we require the decay vertex to be located outside the target
material. We enhance the signal quality by cutting on the isolation variables, $Iso1$ and
$Iso2$. The isolation variable $Iso1$ requires that the tracks forming the $D$ candidate
vertex have a confidence level smaller than the cut to form a vertex with the tracks from the
primary vertex.  The isolation variable $Iso2$ requires that the tracks not assigned 
to the primary or secondary vertices have a confidence level smaller than the cut to form a 
vertex with the $D$ candidate daughters.

{\bf III. $D^+_s \rightarrow K^0_S \pi^-\pi^+\pi^+$ CHANNEL}

For this channel we have excellent secondary vertex resolution with at least three charged 
tracks defining the vertex.  We require $Iso2$ less than 1\% so the secondary 
vertex is isolated from other tracks. We require $Iso1$ less than 1\% to make sure the
$D^+_s$ tracks do not originate at the primary vertex.
The reconstructed mass of the $K_S^0$ must be within four
standard deviations of the nominal $K_S^0$ mass. The typical $K_S^0$ mass resolution
is approximately 6 MeV/$c^2$.
For each pion candidate we require a loose cut that no alternative
hypothesis is greatly favored over the pion hypothesis: $\min{(W_e,W_K,W_p)}-W_\pi > -5$.
For the charged
kaon candidate in the normalization channel we require $W_{\pi} - W_K > 2$. 
We also require the distance $L$ ($\sim$5~mm) between the
primary and secondary vertices divided by its error $\sigma_L$  ($\sim$500~$\micron$)
to be at least 7.
Lastly, we require an additional
$D^{*+} - D^0$ cut for the $K_S^0\pi^-\pi^+\pi^+$ sample. The $K_S^0\pi^-\pi^+\pi^+$ invariant mass 
minus the highest $K_S^0\pi^-\pi^+$ mass combination must be greater than 0.160~GeV/$c^2$.  This
eliminates $D^{*+}$ background events, which simplifies the fitting function.

\begin{figure}[bht]
\begin{center}
\includegraphics[width=6.5cm]{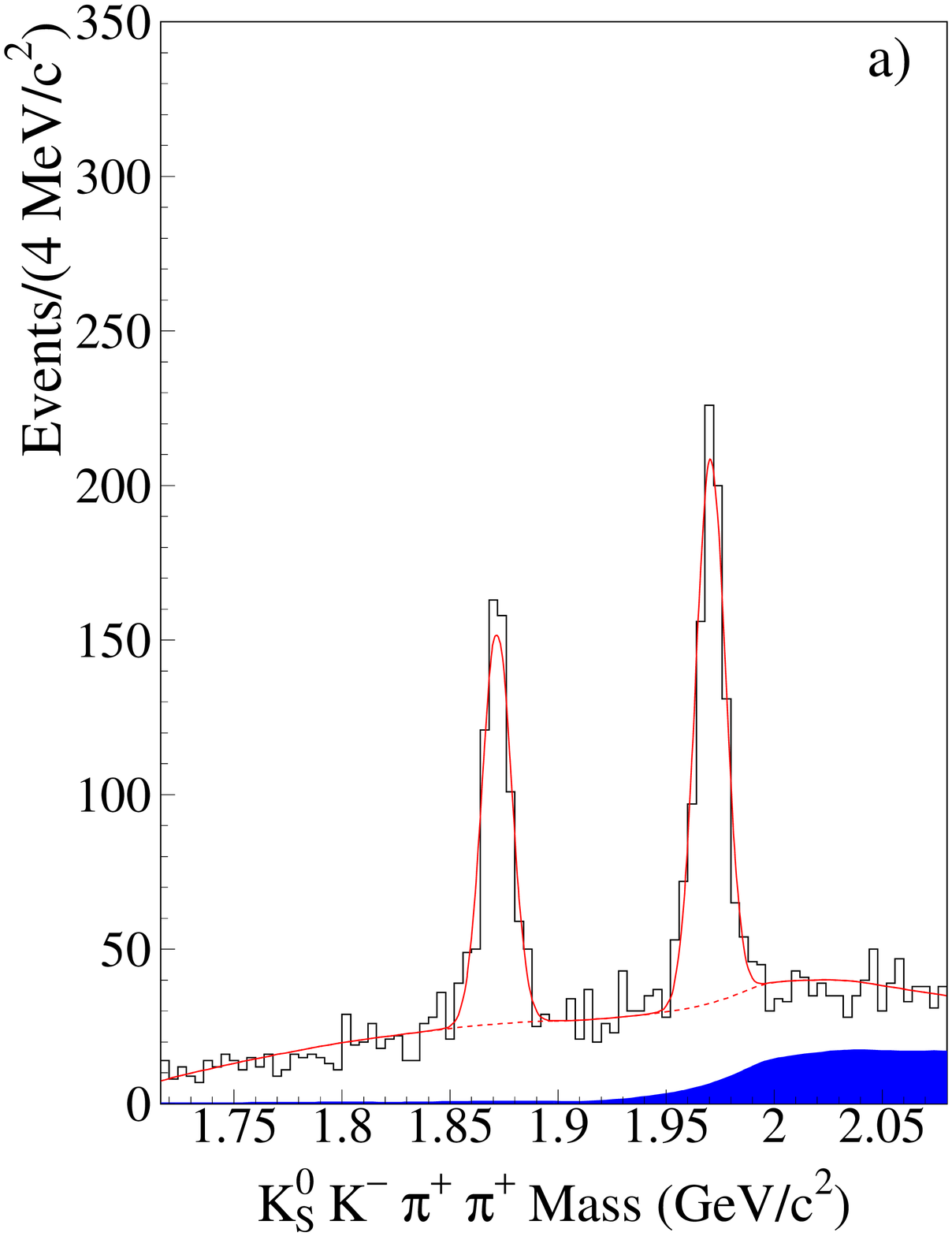}
\includegraphics[width=6.5cm]{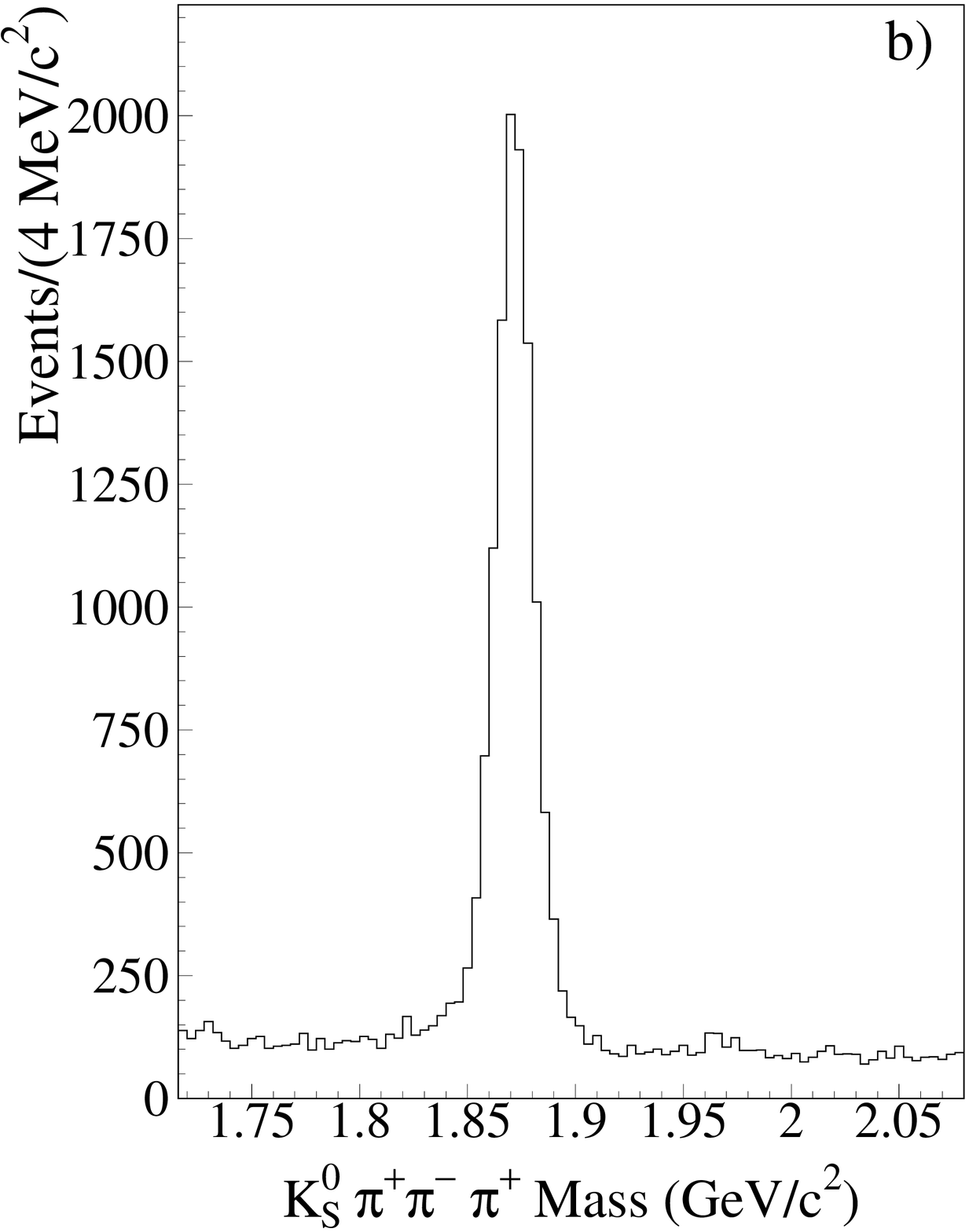}
\includegraphics[width=8cm, height=9cm]{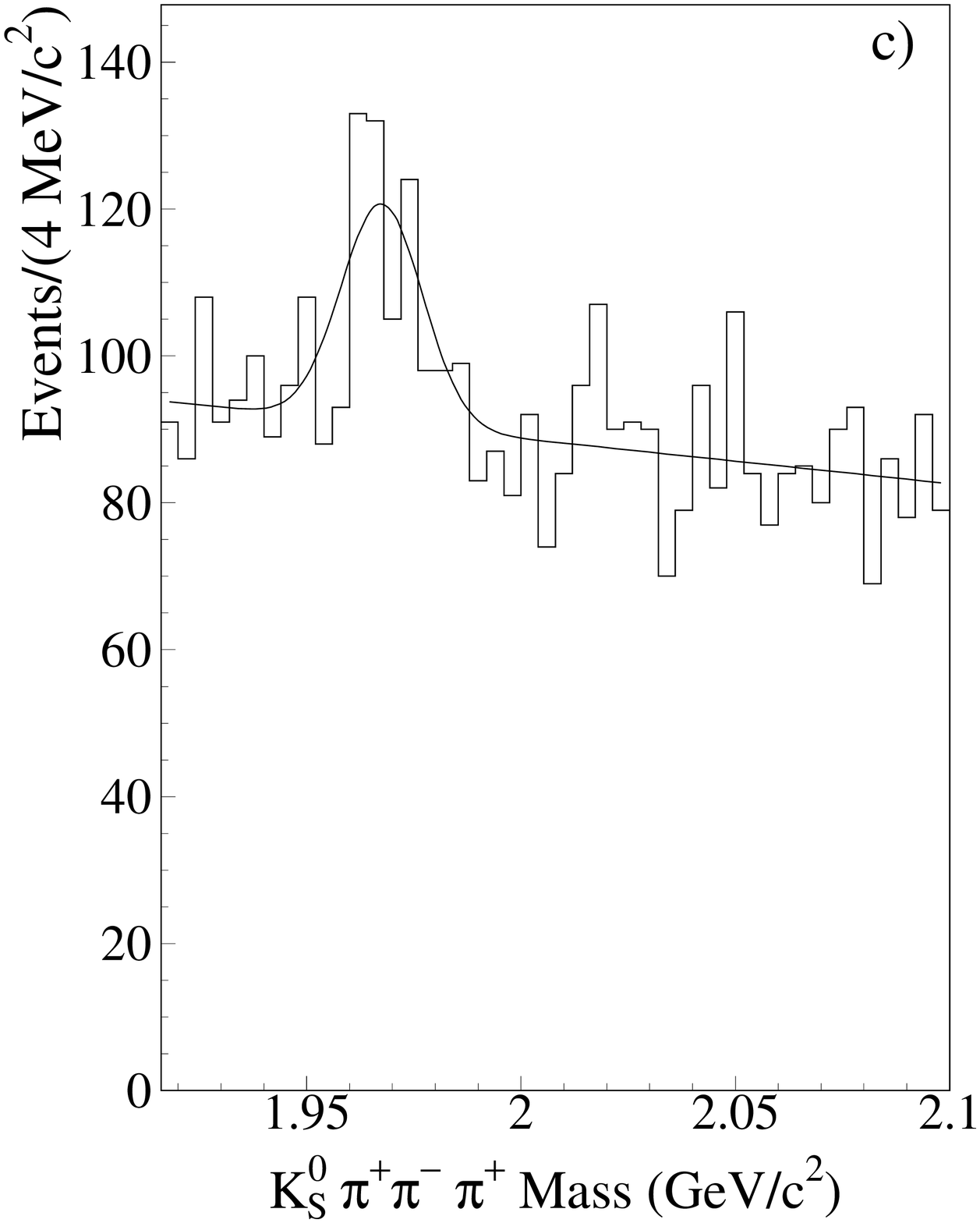}
\end{center}
\caption{Invariant mass distributions for  (a) $K_S^0 K^-\pi^+\pi^+$ (background reflection
from a mismeasured pion from $D^+ \rightarrow K_S^0 \pi^-\pi^+\pi^+$ is included), both the 
$D^+$ and $D^+_s$ signals are evident,
(b) $K_S^0 \pi^-\pi^+\pi^+$ (not fitted to show the large Cabibbo favored $D^+$ contribution),
(c) $K_S^0 \pi^-\pi^+\pi^+$ (with the invariant mass only plotted above the $D^+$ mass).
The mass distribution is fit with a Gaussian with the width fixed from
Monte Carlo for the $D^+_s$ signal and a first degree polynomial for the background.}
\label{fig:ks4h_mass}
\end{figure}

Figure~\ref{fig:ks4h_mass}(a) presents the invariant mass plot for the normalization channel $K_S^0K^-\pi^+\pi^+$ which 
is the cleanest four body $D^+_s$ decay containing a $K^0_S$. The figure contains the Cabibbo suppressed channel
from the $D^+$ as well as the Cabibbo favored $D^+_s$ signal. We fit the $D^+$ and $D^+_s$ signals with
Gaussians. We include a background contribution from $D^+ \rightarrow K_S^0\pi^-\pi^+\pi^+$ where the
$\pi^-$ is misidentified as a kaon and the shape is determined from a Monte Carlo simulation. The combinatoric
background is fit with a 2$^{nd}$ degree  polynomial.  We find $763 \pm 32$ $D^+_s$ signal events at 
$L/\sigma_L > 7$. It is worth noting that this channel has been previously 
studied by the FOCUS
Collaboration and the signal yields reported in this paper are comparable to the results 
already published~\cite{Link:2001r}. 

Figure~\ref{fig:ks4h_mass}(b) shows
the $K_S^0\pi^-\pi^+\pi^+$ invariant mass plot for events that satisfy the above
cuts. The plot is dominated by the Cabibbo favored decay 
$D^+ \rightarrow K_S^0 \pi^-\pi^+\pi^+$
while the $D^+_s$ signal is barely visible. Figure~\ref{fig:ks4h_mass}(c) is the same $K_S^0\pi^-\pi^+\pi^+$ invariant mass
distribution in the region above the $D^+$ peak.
The Figure~\ref{fig:ks4h_mass}(c) mass distribution is fit with a Gaussian with the width fixed from 
Monte Carlo for the signal and a first degree polynomial for the background.    
A signal of $179 \pm 36$ $D_s^+$ events is found from the fit.

\begin{figure}[bht]
\begin{center}
\includegraphics[width=13cm]{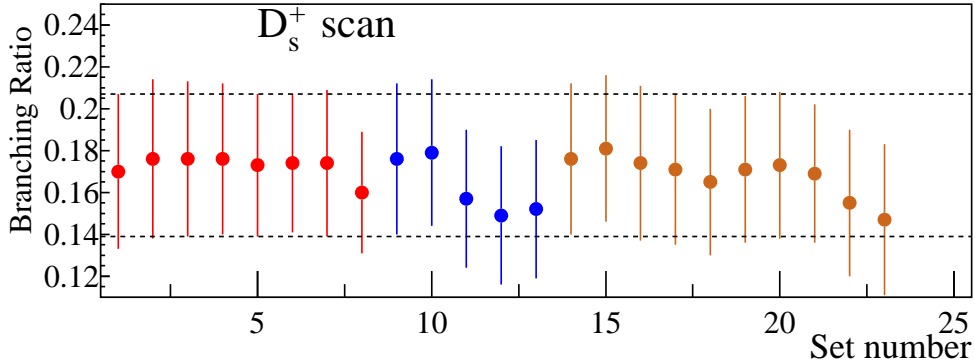}
\end{center}
\caption{The ratio of branching fractions for $D^+_s\rightarrow
K_S^0\pi^-\pi^+\pi^+$ relative to $D^+_s\rightarrow
K_S^0K^-\pi^+\pi^+$ as a function of significance of separation between the
primary and secondary (first eight sets), isolation of the secondary (next five sets),
and confidence level of the
secondary vertex (final 10 sets).}
\label{fig:sys_ks4h}
\end{figure}

We measure the branching fraction of the $D^+_s\rightarrow
K_S^0\pi^-\pi^+\pi^+$ mode relative to $D^+_s\rightarrow
K_S^0K^-\pi^+\pi^+$. The relative efficiency is determined by Monte
Carlo simulation. The relative branching fraction is reported assuming
non-resonant decays for both channels.  
We test for dependency on
cut selection in both modes by individually varying each cut. In Figure~\ref{fig:sys_ks4h}
we present the ratio of branching fractions for $D^+_s\rightarrow
K_S^0\pi^-\pi^+\pi^+$ relative to $D^+_s\rightarrow
K_S^0K^-\pi^+\pi^+$ as a function of significance of separation between the
primary and secondary, isolation of the secondary, and confidence level of the
secondary vertex. 

We studied systematic effects due to uncertainties in the reconstruction
efficiency, in the unknown resonant substructure, and in the fitting procedure.
To determine the systematic error due to the reconstruction efficiency we follow
a procedure based on the S-factor method used by the Particle Data Group~\cite{PDG:2006ed}.
For each mode we split the data sample into two independent subsamples based on
$D^+_s$ momentum,  particle versus antiparticle, decays
inside the target material versus outside of target material, and on the period of 
time in which the data was collected. These
splits provide a check on the Monte Carlo simulation of charm production, on the
vertex detector (which was upgraded during the run), and on the simulation
of the detector stability. We then define the split sample variance as the
difference between the scaled variance and the statistical variance if the
former exceeds the latter. The method is described in detail in reference~\cite{Link:2003al}.
We vary the subresonant states
in the Monte Carlo and use the variance in the branching ratios as a
contribution to the systematic error. We investigate the systematic effects
based on different fitting procedures and we find this contribution to be small.
The branching ratio is evaluated under
various cut selection criteria, and the variance of the results is used as an
additional systematic error.
The systematic effects are then all added in
quadrature to obtain the final systematic error. Table~\ref{tab:sys_ks4h} summarizes the
contributions to the systematic errors for the $\frac{\Gamma(D^+_s\!\rightarrow
K_S^0\pi^-\pi^+\pi^+)}{\Gamma(D^+_s\!\rightarrow
K_S^0K^-\pi^+\pi^+)}$ branching ratio.  The result, $\frac{\Gamma(D^+_s\!\rightarrow
K_S^0\pi^-\pi^+\pi^+)}{\Gamma(D^+_s\!\rightarrow
K_S^0K^-\pi^+\pi^+)} = 0.18 \pm 0.04 \pm 0.05$, is summarized in Table~\ref{tab:results}.

\begin{table}[h]
\begin{center}
\caption{Summary of the systematic error contributions for $D^+_s\rightarrow
K_S^0\pi^-\pi^+\pi^+$.\label{tab:sys_ks4h}}
\begin{tabular}{ccc} \hline \hline
Contribution &  $\frac{\Gamma(D^+_s\!\rightarrow
K_S^0\pi^-\pi^+\pi^+)}{\Gamma(D^+_s\!\rightarrow
K_S^0K^-\pi^+\pi^+)}$\\
\hline
D Momentum \& Run Period & 0.04\\
Split Target (in versus out) & 0.03 \\
Set of Cuts Selection     & 0.01 \\
Fit Variance             & 0.01 \\
Resonant Substructure    & 0.01\\
\hline
Total                    & 0.05\\
\hline
\hline
\end{tabular}
\end{center}
\end{table}

{\bf IV. $D^+_s \rightarrow K^0_S \pi^+$ CHANNEL}

This is a challenging channel to reconstruct as we typically only have a detached
silicon track from the production vertex and a $K^0_S$ to indicate a candidate.
Several criteria are used to improve the signal over background. Since any signal was
expected to be small the selection criteria are optimized using Monte Carlo signal events
and sideband background events. The figure of merit used was $S/\sqrt{B}$ and the cuts
were chosen sequentially. At each step, the $S/\sqrt{B}$ distribution was determined for the
full range of each cut. The cut which had the highest $S/\sqrt{B}$ was selected and a
cut was made more conservative than the maximum $S/\sqrt{B}$ point. The procedure was then
repeated until no further improvement was possible.
 
For the 90\% of the $K^0_S$ decays that occur after the $K^0_S$ has passed through  
the silicon strip detector, we employ a specialized vertex algorithm to 
locate the $K^0_S\pi^+$ vertex.  We use the momentum information from the 
$K^0_S$ decay and the silicon track of the pion to form a candidate $D^+_s$ vector.
This vector is intersected with candidate production vertices which are formed with 
two other silicon tracks. When the $D$ vector is forced to originate at the production 
vertex, we can compute a confidence level that the $D^+_s$ vector formed a vertex 
with the 
charged daughter.  As the type and resolution of $K^0_S$ is integral to finding the
$D^+_s$ vertex, the significance of separation, $L/\sigma_L$,  between the production
and $D^+_s$ decay vertices were varied according to the $K^0_S$ decay type. The
$L/\sigma_L$ cuts varied from 7--11.  This mode also required $Iso2 < 2\%$. 

The normalization channel is the Cabibbo favored $D^+_s \rightarrow K^0_S K^+$.  The selection
criteria for this channel (with the exception of particle identification) are identical
to $D^+_s \rightarrow K^0_S \pi^+$. The momentum of the $D_s^+$ and the charged hadron in the
$D_s^+$ decay must be greater than 45~GeV/$c$ and 12~GeV/$c$, respectively. 
To reduce the effect of long-lived decays and reinteractions, the proper decay time must be
less than $2.5$~ps with an uncertainty less than $0.12$~ps.
To help separate charm from combinatoric background, 
a momentum asymmetry cut on the two body $D^+_s$ decay was used: 
$\left|\frac{p(K^0_S) - p(h^+)}{p(K^0_S) + p(h^+)}\right| < 0.75$.

For the $K^+$ candidate the negative log-likelihood kaon hypothesis, $W_K =
-2$ ln(kaon likelihood) must be favored over the corresponding pion hypothesis
$W_{\pi}$ by $W_{\pi} - W_K > 4$ while for the signal mode, the $\pi^+$ candidate must
have $W_K - W_{\pi} > -1$.  The first cut serves to dramatically reduce the potentially
large $D^+\to K_S^0\pi^+$ background which peaks at the $D_s^+$ mass when reconstructed as
$K_S^0K^+$ while the second cut reduces $D_{(s)}^+\to K_S^0K^+$ background which is smaller
to begin with and peaks below the $D_s^+$ mass when reconstructed as $D_s^+\to K_S^0\pi^+$.

Fitting the $D_s^+\!\to\!K_S^0\pi^+$ mass plot is complicated by the presence of the 
large $D^+\!\to\!K_S^0\pi^+$ signal.
Since the resolution of the state is relatively poor ($\sigma \approx 13$~MeV/$c^2$) 
there is very little space between the $D^+$ and $D_s^+$
peaks to estimate the background.  The fit used to obtain the central value has five contributions.  The first
contribution is the $D^+\!\to\!K_S^0\pi^+$ signal which is fit with a distribution obtained from smoothing a
Monte Carlo sample of reconstructed $D^+\!\to\!K_S^0\pi^+$ events.  The mean and yield are fitted parameters.
The second contribution is the $D_s^+\!\to\!K_S^0\pi^+$ signal which is also fit with a distribution obtained from smoothing a
Monte Carlo sample of reconstructed $D_s^+\!\to\!K_S^0\pi^+$ events.  In this case, the mean is fixed.
The third and fourth contributions are reflections from
$D_s^+\!\to\!K_S^0K^+$ and $D^+\!\to\!K_S^0K^+$.  The reflection shapes are obtained from Monte Carlo 
samples of generated
$D_{(s)}^+\!\to\!K_S^0K^+$ events reconstructed as $D_s^+\!\to\!K_S^0\pi^+$.  The level 
is found by taking the same generated
events, reconstructing them properly, and determining the yield.  This Monte Carlo yield 
is then compared to the yield of the
data $D_s^+\!\to\!K_S^0K^+$ and $D^+\!\to\!K_S^0K^+$ and this factor multiplies the reflection shapes.  
Finally, the fifth contribution is a quadratic polynomial to account for generic combinatorial 
background.

The $K_S^0K^+$ mass plot is also fit with five contributions.  The  $D_s^+\!\to\!K_S^0K^+$ and 
$D^+\!\to\!K_S^0K^+$ are fit with
functions obtained from smoothing reconstructed Monte Carlo samples.  The masses and yields are fitted in both cases.  The
reflection from $D^+\!\to\!K_S^0\pi^+$ is also obtained from Monte Carlo and fixed based on the 
number of reconstructed
$D^+\!\to\!K_S^0\pi^+$ events in data.  The fourth contribution, a reflection from $D_s^+\!\to\!K_S^0K^+\pi^0$ is 
allowed in the fit.
The shape is obtained from Monte Carlo simulation but the level is allowed to vary in the fit since the branching ratio is
poorly known and we do not have a fully reconstructed sample available. As before, the fifth
contribution is generic combinatoric background which is modeled with a quadratic polynomial.
                                                                                 
\begin{figure}[bht]
\begin{center}
\includegraphics[width=6.5cm]{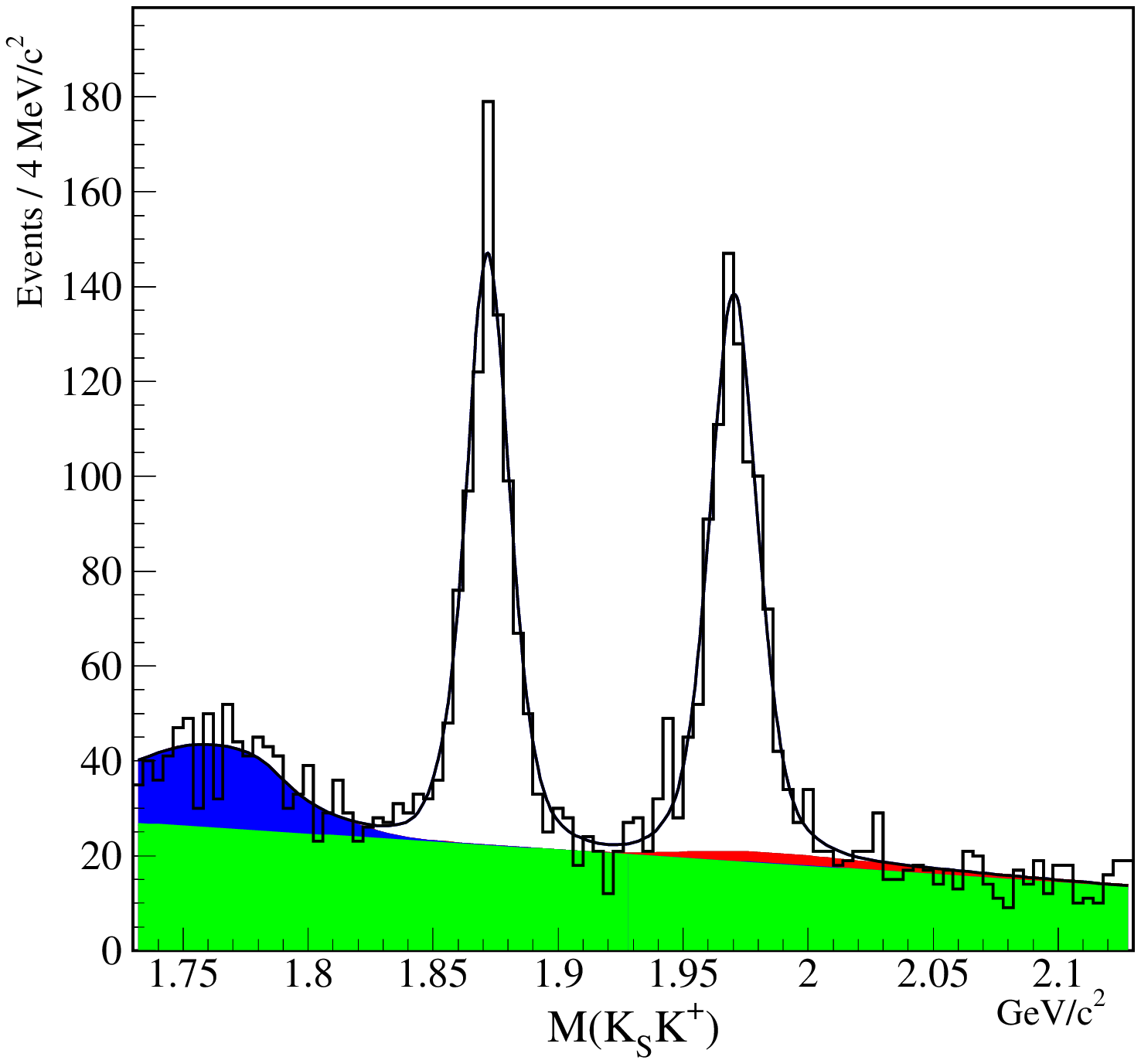}
\includegraphics[width=6.5cm]{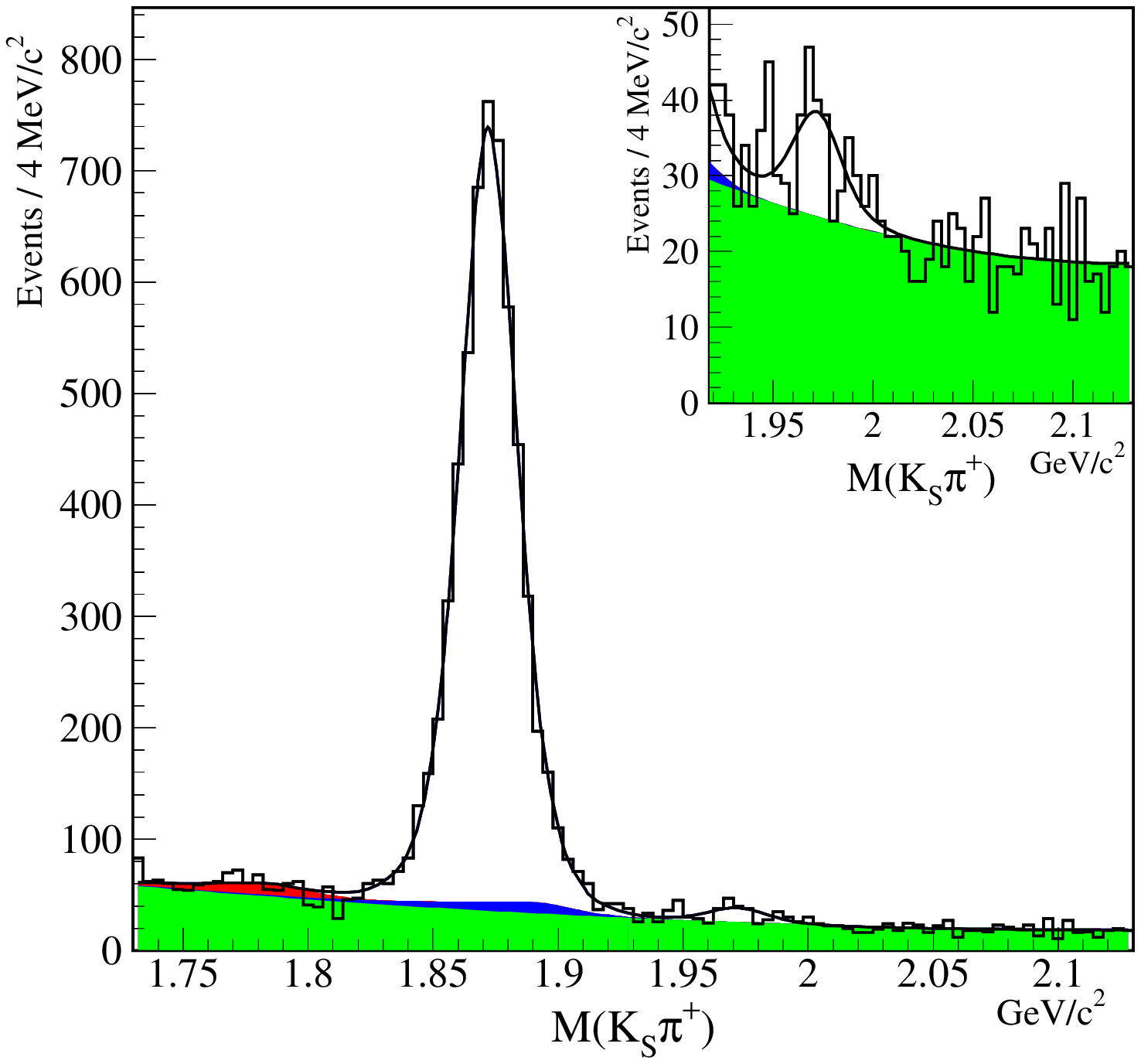}
\end{center}
\caption{Invariant mass distributions for $K_S^0 K^+$ (left) and
$K_S^0 \pi^+$ (right).
The fits are over the entire mass range.  Most of the background is 
modeled by a quadratic polynomial.  The remaining background is due to 
reflections and is a different shade.  The $K_S^0K^+$ mode has a large 
reflection component from $D_s^+\!\to\!K_s^0K^+\pi^0$ below the $D^+$ peak 
and a small reflection component from $D^+\!\to\!K_s^0\pi^+$ under the 
$D_s^+$ peak.  The $K_s^0\pi^+$ has small reflection contributions below 
(under) the $D^+$ peak from $K_S^0\pi^+$ decays from $D^+$ $(D_s^+)$.  All 
signal and reflection shapes come from a Monte Carlo simulation.}
\label{fig:ks1h_mass}
\end{figure}

From the $K_S^0\pi^+$ fit shown in Fig.~\ref{fig:ks1h_mass} we obtain a $D_s^+$ yield of $113 \pm 26$ events.
The $K_S^0K^+$ fit presented in Fig.~\ref{fig:ks1h_mass} gives a yield of $777 \pm 36$ $D_s^+$ events and the
number of events found for the $D_s^+\!\to\!K_S^0K^+\pi^0$
reflection is consistent with PDG branching ratios and our efficiency.
                                 
The systematic uncertainties are divided into cut variants and fit variants.  
In both cases the systematic
uncertainty is obtained from the square root of the standard deviation of the values weighted by the individual
uncertainty.  The actual procedure is as follows.  For each variant (but not the default), 
the branching ratio $B\!R_i$
is calculated along with the uncertainty $\sigma_i$.  The average, weighted by the inverse of 
the square of the uncertainty,
is calculated
\begin{equation}
\overline{B\!R} \;=\; \frac{\sum_i \frac{B\!R_i}{\sigma_i^2}}{\sum_i \frac{1}{\sigma_i^2}}.
\end{equation}
The systematic uncertainty is obtained from the square root of the standard deviation which comes from a ``weighted'' $\chi^2$:
\begin{equation}
\sigma_{\textrm{sys}} \;=\; \sqrt{\frac{\sum_{i=1}^N \left( \sigma_0^2 \frac{B\!R_i - \overline{B\!R}}{\sigma_i^2}\right)^2}{N-1}}
\end{equation}
where $\sigma_0$ is the uncertainty on the default measurement.
                                                                                                                  
For each of the cut variants,
both the $D_s^+\!\to\!K_S^0\pi^+$ and $D_s^+\!\to\!K_S^0K^+$ samples are changed the same (with the exception
of particle identification cuts).  The variations are consistent with statistical fluctuations and the systematic
uncertainty is determined from the standard deviation which is dominated by the $D_s^+\!\to\!K_S^0\pi^+$
variations. The systematic uncertainty from the cut variant is $\sigma^{cut}_{sys} = 0.010$.  

The systematic uncertainty in estimating the yield of $D_s^+\!\to\!K_S^0K^+$ events is negligible 
compared to estimating the
yield of $D_s^+\!\to\!K_S^0\pi^+$ events.  Therefore, for the fit variants we vary how  the $K_S^0\pi^+$ mass 
plot is fitted.  Some of the variations include fitting with a Gaussian, 
allowing the mass and width to float, and fitting only above the $D^+$ mass peak.
The variation in the $D_s^+\!\to\!K_S^0\pi^+$ yield, again weighted by the uncertainty squared, 
gives the systematic uncertainty. The systematic uncertainty on the yield from the fit variations is 
9.0 events which corresponds to a relative uncertainty of 8.0\% and translates into a systematic 
uncertainty on the branching ratio of $\sigma^{fit}_{sys} = 0.008$.  Adding the cut and fit systematic
uncertainties in quadrature gives a total systematic uncertainty on the branching ratio of 0.013.
                                                                                                                  
{\bf V. SUMMARY OF RESULTS}

In conclusion we have presented the
first evidence of the Cabibbo suppressed decay mode $D_s^+\rightarrow K_S^0 \pi^-\pi^+\pi^+$ and 
measured the relative branching ratio of 
{$\frac{\Gamma(D_s^+\rightarrow
K_S^0\pi^-\pi^+\pi^+)}{\Gamma(D_s^+\rightarrow
K_S^0K^-\pi^+\pi^+)}$} = $0.18 \pm 0.04 \pm 0.05$. A naive expectation for this
 branching ratio is tan$^2\theta_C =0.054$.  Compared with this expectation the
branching ratio is more than 3 times larger.  One contributing factor is there 
is more phase space 
available in the $D^+_s \rightarrow K^0_S\pi^-\pi^+\pi^+$ decay than in the
$D^+_s \rightarrow K^0_S K^-\pi^+\pi^+$ decay. Another factor is that the $K^0_S$ 
in the denominator of the ratio comes from a $\overline{K^0}$. In the numerator the 
$K^0_S$ may be the result of either a $\overline{K^0}$ or a ${K^0}$ decay. Perhaps a
better understanding of this ratio would result from reporting the ratio
{$\frac{\Gamma(D_s^+\rightarrow
K_S^0\pi^-\pi^+\pi^+)}{\Gamma(D_s^+\rightarrow
K_S^0K^-\pi^+\pi^+) + \Gamma(D_s^+\rightarrow
K_S^0K^+\pi^+\pi^-)}$}.  Using the branching ratio reported in reference~\cite{Link:2001r} for
$\frac{\Gamma(D_s^+\rightarrow
K_S^0K^+\pi^+\pi^-)}{\Gamma(D_s^+\rightarrow
K_S^0K^-\pi^+\pi^+)} = 0.586 \pm 0.052 \pm 0.043$ we find
{$\frac{\Gamma(D_s^+\rightarrow
K_S^0\pi^-\pi^+\pi^+)}{\Gamma(D_s^+\rightarrow
K_S^0K^-\pi^+\pi^+) + \Gamma(D_s^+\rightarrow
K_S^0K^+\pi^+\pi^-)} \approx 0.11$}. 
 
We also present evidence for  $D_s^+\rightarrow K_S^0 \pi^+$ 
 and measure its branching fraction relative to $D_s^+\rightarrow K_S^0 K^+$:
{$\frac{\Gamma(D_s^+\rightarrow K_S^0\pi^+)}{\Gamma(D_s^+\rightarrow
K_S^0K^+)}$} = $0.104 \pm 0.024 \pm 0.013$. This branching ratio is also larger than
$\tan^2\theta_C$, but is slightly smaller than predictions~\cite{Verma:1991a,Buccella:1995a,Buccella:1996a}
which range from 14\% to 17\%.
The results are summarized in Table~\ref{tab:results}.

\begin{table}[htb]
\begin{center}
\caption{Branching ratios, event yields,  and efficiency ratios for modes
involving a $K_S^0$. All
branching ratios are inclusive of subresonant modes.\label{tab:results}}
\begin{tabular}{cccc} \hline \hline
Decay Mode&Ratio of Events&Efficiency Ratio&Branching Ratio\\
\hline
$\frac{\Gamma(D_s^+\!\rightarrow K_S^0\pi^-\pi^+\pi^+)}{\Gamma(D_s^+\!\rightarrow
K_S^0 K^-\pi^+\pi^+)}$&$\frac{179 \pm 36}{763 \pm 32}$&1.34&$0.18 \pm 0.04 \pm 0.05$\\
$\frac{\Gamma(D^+_s\!\rightarrow K_S^0 \pi^+)}{\Gamma(D^+_s\!\rightarrow
K_S^0 K^+)}$&$\frac{113 \pm 26}{777 \pm 36}$&1.39&$0.104 \pm 0.024 \pm 0.013$\\
\hline
\hline
\end{tabular}
\end{center}
\end{table}

{\bf VI. ACKNOWLEDGEMENTS}

We acknowledge the assistance of the staffs of Fermi National
Accelerator Laboratory, the INFN of Italy, and the physics departments
of
the
collaborating institutions. This research was supported in part by the
U.~S.
National Science Foundation, the U.~S. Department of Energy, the Italian
Istituto Nazionale di Fisica Nucleare and
Ministero della Istruzione, Universit\`a e
Ricerca, the Brazilian Conselho Nacional de
Desenvolvimento Cient\'{\i}fico e Tecnol\'ogico, CONACyT-M\'exico, and
the Korea Research Foundation of
the Korean Ministry of Education.

\bibliographystyle{apsrev}
\bibliography{ds_plb}

\end{document}